\documentclass[a4paper,11pt]{article}
\usepackage{pos}
\usepackage[utf8]{inputenc}
\usepackage{physics}
\usepackage{amsmath}
\usepackage{graphicx}
\usepackage[margin=0.1in]{subcaption}
\usepackage{siunitx}
\usepackage{hyperref}
\usepackage{cleveref}
\usepackage{floatrow}
\newcommand{\brho}{\boldsymbol{\rho}}

\newcommand{\F}[1]{\mathcal{F}\left\{ #1 \right\}}

\title{Mass estimates of the SU(2) $0^{++}$ glueball from spectral methods}

\author[a,b]{David Dudal}
\author*[c]{Orlando Oliveira}
\author[a]{Martin Roelfs}

\affiliation[a]{Department of Physics, KU Leuven Campus Kortrijk--Kulak, Etienne Sabbelaan 53/7657, Kortrijk, Belgium}
\affiliation[b]{Department of Physics and Astronomy, Ghent University, Krijgslaan 281-S9, Gent, Belgium}
\affiliation[c]{CFisUC, Department of Physics, University of Coimbra, Coimbra, Portugal}

\emailAdd{david.dudal@kuleuven.be}
\emailAdd{orlando@uc.pt}
\emailAdd{martin.roelfs@kuleuven.be}

\abstract{The estimation of the Källén-Lehmann spectral density from gauge invariant lattice QCD two point correlation functions is proposed, and explored via an inversion strategy based on Tikhonov regularisation. We test the method on a mesonic toy model, showing that our methodology is competitive with the traditional Maximum Entropy Method. As proof of concept the SU(2) glueball spectrum for the quantum numbers $J^{PC}=0^{++}$ is investigated, for various values of the lattice spacing, using the published data of Phys.\ Rev.\ D 102 (2020) 5, 054507. Our estimates for the ground state mass are in good agreement with the traditional approach, which is based on the large time exponential behaviour of the correlation functions. Furthermore, the spectral density also contains hints of excites states in the spectrum. Spectroscopic analysis of glueball two-point functions therefore provides a straightforward and insightful alternative to the traditional method based on the large time exponential behaviour of the correlation functions.}

\FullConference{%
 The 38th International Symposium on Lattice Field Theory, LATTICE2021
  26th-30th July, 2021
  Zoom/Gather@Massachusetts Institute of Technology
}

\begin{document}
    \maketitle
	\section{Motivation}
	
	A first principles approach to access the hadron spectra, which has been advancing since the beginning of the eighties from last century \cite{Ishikawa:1982tb},
	is lattice QCD. In a typical lattice QCD computation of the bound state mass, an appropriate two point correlation function is evaluated and, from its large time decay behaviour and the associated slope, the ground state mass is estimated \cite{Montvay:1994cy}.
		
	We discuss an alternative way of accessing the particle masses: via the computation of the K\"all\'en-Lehmann spectral
	representation associated with the momentum space particle propagator \cite{Peskin:1995ev}.
	A possible advantage of using spectral representations compared to a conventional lattice calculation is that it does not necessarily require the use of smearing, or other techniques, to improve the Monte Carlo signal to noise ratio, see e.g.
	\cite{Blossier:2009kd,Athenodorou:2020ani} and references therein.
	
	In the current work we focus on the SU(2) glueball with quantum numbers  $J^{PC} = 0^{++}$,  i.e.~the scalar glueball. More details can be found in the preprint \cite{Dudal:2021gif} on which this talk was based.

\section{Setup}
	Let us describe our procedure to access the K\"all\'en-Lehmann spectral representation $\rho(\omega)$ from a two point correlation function $G(p)$.
	The relation between these two functions reads
		\begin{equation}
		G(p) = \int_{0}^{\infty} \frac{2 \omega \rho(\omega) \dd{\omega}}{\omega^2 + p^2} \label{eq:propagator_gl}
		= \int_{-\infty}^{\infty} \frac{\rho(\omega) \dd{\omega}}{\omega - i p}  \ .\notag
	\end{equation}
Depending on the dimensionality of the involved quantum operators, proper subtractions are needed to guarantee finite results \cite{Weinberg:1995mt}. This corresponds to adding appropriate contact counterterms, or polynomials in momentum space. In general, one has
		\begin{align}
		G(p^2)
		&= \sum_{k=0}^{n-1} a_{k} (p^2-\bar{p}^2)^{k} +  (- p^2 + \bar{p}^2)^{n} \int_{0}^{\infty} \frac{2 \omega \tilde{\rho}(\omega) \dd{\omega}}{\omega^2 + p^2} \label{eq:expanded_propagator}\\
		&=  \sum_{k=0}^{n-1} a_{k} (p^2-\bar{p}^2)^{k} +  (- p^2 + \bar{p}^2)^{n} \int_{-\infty}^{\infty} \frac{\tilde{\rho}(\omega) \dd{\omega}}{\omega - ip} \notag
	\end{align}
	with
		\begin{align}
		a_{n} &= \frac{1}{n!} \pdv[n]{G(p^2)}{(p^2)}\Bigg\vert_{p^2=\bar{p}^2}, \label{eq:taylor_coeff}\qquad
		\tilde{\rho}(\omega)~=~\frac{\rho(\omega)}{(\omega^2 + \bar{p}^2)^n},
	\end{align}
	and $\bar{p}^2$ is a reference momentum scale at which the subtraction is done. Unfortunately, performing subtractions on numerical data is extremely sensitive to the choice of $\bar{p}$ and results in relative rapid 	variation of $\{ a_k \}$ with the subtraction point \cite{Oliveira:2012eu}.  An elegant way to perform the subtractions, without actually having to do these, is by considering the Fourier transform of $G(p)$ and look at the Schwinger function defined as
		\begin{equation}
		C(\tau) = \F{G(p)}(\tau) = \int_{-\infty}^{\infty} G(p^2) \bigg\vert_{\vec{p} = 0, p_4 \neq 0} ~ e^{-i p_4 \tau} \dd{p_4}.
		\label{Eq:Schwinger}
	\end{equation}
	For $\bar{p} = 0$, we get
		\begin{align}
		C(\tau) &= \F{G(p)}(\tau) \\
		&\hspace{-0.7cm}= \mathcal{L}\left\{ \rho(\omega) \right\}(|\tau|) + 2 \pi \sum_{k=0}^{n-1} a_{k} (-1)^k \delta^{(2k)}(\tau)  + 4 \pi (-1)^{n+1} \sum_{k=2}^{n} \delta^{(2(n-k)+2)}(\tau) \int_{0}^{\infty} \dd{\omega} \omega^{2k-3} \tilde{\rho}(\omega). \notag
	\end{align}
	The important observation however, is that $C(\tau) = \mathcal{L}\left\{ \rho(\omega) \right\}(|\tau|)$ when $\tau~\neq~0$, and equal to a sum of (derivatives of) Dirac delta functions when $\tau = 0$.
	Therefore, $\rho(\omega)$ can be recovered by taking $C(\tau)$ for $\tau>0$, and inverting the Laplace transformation.
	
	Because the inverse Laplace transform is an ill-defined numerical problem, regularization is necessary in order to perform the
	inversion. We shall use Tikhonov regularization, similar to \cite{Dudal:2013yva,Dudal:2019gvn}. However,
	because glueballs are observable particles, their spectral density $\rho(\omega)$ is non-negative. Therefore we shall implement Tikhonov
	regularization using Non-Negative Least Squares (NNLS) \cite{Lawson1995}, to ensure
	a positive spectral function $\rho(\omega) \geq 0$.

 The most widely used approach for spectral function extraction from numerical data is based on the maximum entropy method \cite{Asakawa:2000tr}. We will compare our outcome with the latter.

	\section{The Numerical Method}\label{sec:method}
	
	In a lattice simulation the propagator $G(p_n)$
	is computed on a finite set of evenly spaced momenta $p_n$.
	Given a data set $\{ G(p_n) \} := \{ G(p_0), \ldots, G(p_{N-1}) \}$, the Schwinger function is computed using DFFT, resulting in
	a data set $\{ C(\tau_k) \}$, where
		\[ C(\tau_k) = \sum_{n=0}^{N-1} G(p_n) e^{- i 2 \pi k n / N}. \]
	The Laplace transformation to access the spectral function
		\begin{align}
		C(\tau_k) = \mathcal{L}\left\{ \rho(\omega) \right\}(\tau_k) = \int_0^\infty e^{- \omega \tau_k} \rho(\omega) \dd{\omega},
	\end{align}
	can be written as a matrix equation
		\begin{equation*}
		\boldsymbol{C} = \boldsymbol{K} \brho,
	\end{equation*}
	with the elements of $\boldsymbol{K}$ defined as
		\begin{equation*}
		\boldsymbol{K}_{k\ell} :=  e^{- \omega_\ell \tau_k} \Delta \omega.
	\end{equation*}
	Since $\brho$ needs to be obtained, and a direct solution is impossible due to the near zero singular values of $\mathbf{K}$, the original
	problem is replaced by the minimisation of the Tikhonov regularizing functional
		\begin{equation}
		J_\alpha = \norm{\boldsymbol{K} \brho - \boldsymbol{C}}_2^2 + \alpha^2 \norm{\brho - \brho^*}_2^2
		\label{eq:tikhonov}
	\end{equation}
	where $\alpha^2 > 0$ is the Tikhonov parameter and $\brho^*$ is a prior estimate for $\brho$. In order to impose the constraint $\brho \geq 0$, define
		\begin{align}
		\boldsymbol{A} = \mqty(\boldsymbol{K} \\ \alpha \mathbb{1} ),\quad \boldsymbol{b} = \mqty(\boldsymbol{C} \\ \alpha \brho^*),
	\end{align}
	and rewrite $J_\alpha$ as
		\begin{equation}
		J_\alpha = \norm{\boldsymbol{A} \brho - \boldsymbol{b}}_2^2.
		\label{eq:tikhonov_lsq}
	\end{equation}
	The formulation of the problem
	using Eq. \eqref{eq:tikhonov_lsq} can be solved with a non-negative least squares (NNLS) solver such that $\brho \geq 0$ is guaranteed \cite{Lawson1995}.

The regularization parameter $\alpha$, in essence, provides a soft threshold to the singular values of $\boldsymbol{K}$, such that the smallest singular values no longer cause numerical issues. 	Choosing $\alpha$ is a delicate affair, since setting it too small means the problem remains ill-defined, whereas setting it too large destroys a lot of the information contained in the data. We rely on the Morozov discrepancy principle \cite{Dudal:2019gvn}, which states that $\alpha^2$ should be chosen such that
		\begin{equation}
		\norm{\boldsymbol{K} \brho - \boldsymbol{C}}_2^2 = \sum_i \sigma_i^2,
		\label{eq:morozov_gl}
	\end{equation}
	where $\sum_i \sigma_i^2$ is the total variance in the data. The $\alpha^2$ obeying \cref{eq:morozov_gl} is guaranteed to be unique \cite{Kirsch:1996:IMT:236740}, and means that the quality of the reconstruction is identical to the quality of the data.
	
	We sample $\omega$ evenly in logarithmic space from $[10^{-5}, 10^5]$ \SI{}{\GeV} in $N_\omega$ steps.

	\subsection{Test: Meson toy-model}\label{sec:results_meson}
	
	\begin{figure}[hbt]
		\centering
		\includegraphics[width=\textwidth,trim={0.5in 0.4in 0.5in 0.5in},clip]{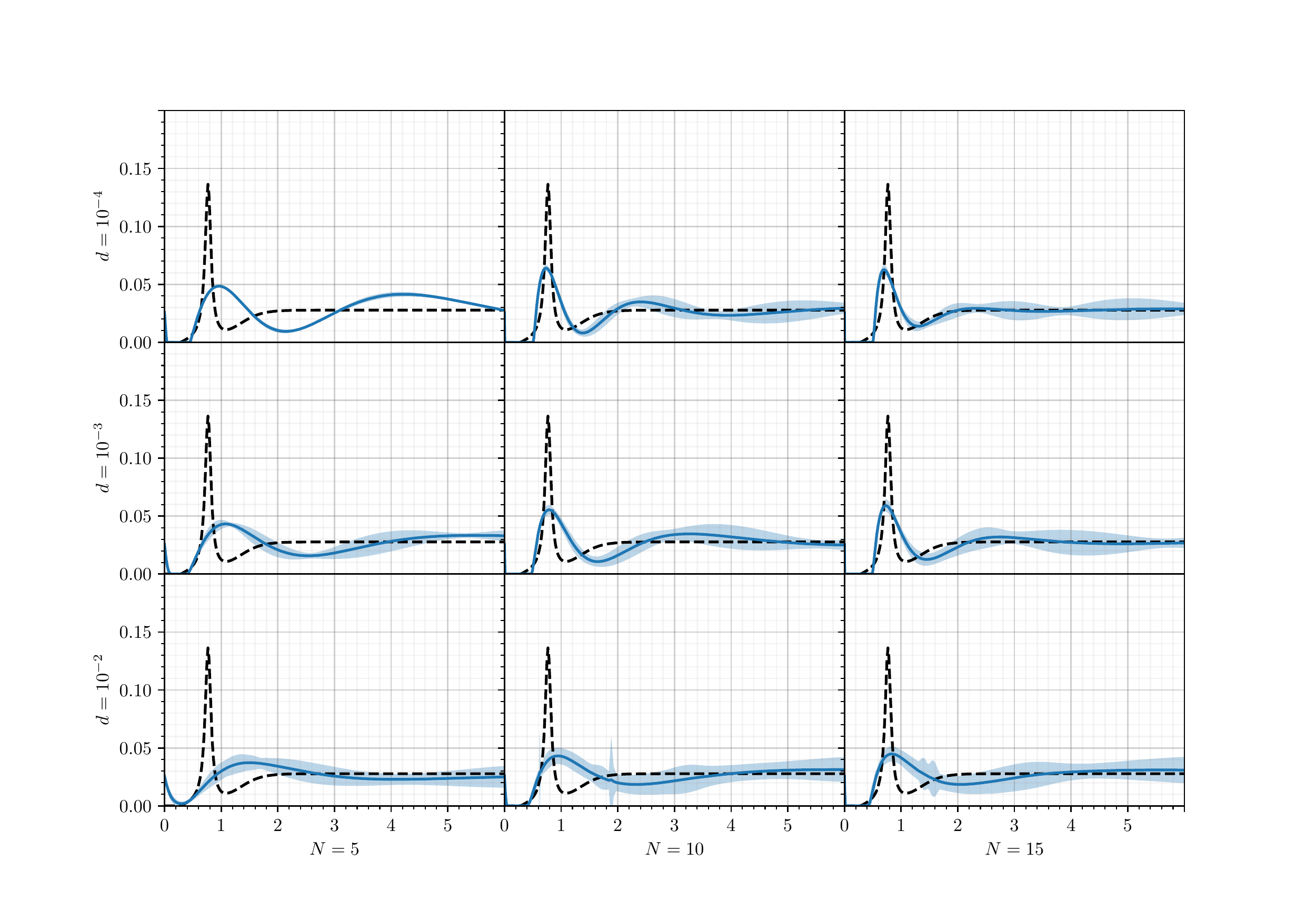}
		\caption{Reconstruction the toy-model spectral density function for various $d$ and $N$. The dashed black line is the original spectral function, while the blue full curve is the reconstructed spectral function as given by the Tikhonov regularized NNLS method.}
		\label{fig:toymodel}
	\end{figure}
	
	In order to investigate the reliability of the method we consider a realistic toy-model, based on a vector-meson model decay into hadrons,
	as used before in \cite{Asakawa:2000tr}. This particular model needs  a single subtraction, and therefore provides an excellent test of the method.
	To allow comparison of our results to those of \cite{Asakawa:2000tr}, we shall use the exact same process to generate the mock data.
	The meson spectral density function is given by
	
	\begin{equation}
		\rho(\omega) = \frac{2}{\pi}  \bqty{F^2_{\rho} \frac{\Gamma_\rho m_\rho }{(\omega^2 - m_\rho^2)^2 + \Gamma_\rho^2 m_\rho^2} + \frac{1}{8 \pi} \pqty{1 + \frac{\alpha_s}{\pi}} \frac{1}{1 + e^{(\omega_0 - \omega) / \delta}}},
	\end{equation}
	with an energy-dependent width
	
	\begin{equation}
		\Gamma_\rho(\omega) = \frac{g_{\rho\pi\pi}^2}{48 \pi} m_\rho \pqty{1 - \frac{4 m_\pi^2}{\omega^2}}^{3/2} \theta\pqty{\omega - 2 m_\pi}.
	\end{equation}
	The empirical values of the parameters are
		\begin{equation}
		m_\rho = 0.77 \text{ GeV}\,, m_\pi = 0.14 \text{ GeV}\,,
		g_{\rho\pi\pi} = 5.45\,, F_\rho = \frac{m_\rho}{g_{\rho\pi\pi}}\,,
		\omega_0 = 1.3 \text{ GeV}\,,\delta = 0.2 \text{ GeV}.
	\end{equation}
    As $\omega \to \infty$, this model behaves like $\rho(\omega \to \infty) = (1 / 4 \pi^2) (1 + \alpha_s/\pi)$. Therefore the integral \cref{eq:propagator_gl} does not converge, and a single subtraction has to be performed, that means $n=1$ in the notation used in \cref{eq:taylor_coeff}.

    Assuming $\alpha_s = 0.3$, the value $(1 / 4 \pi^2) (1 + \alpha_s/\pi) = 0.0277$ can be used as the prior, but identical to \cite{Asakawa:2000tr} we shall use the slightly smaller value $\rho_\text{prior}~=~0.0257$.
    In order to generate mock data, we compute $C_\text{orig}(\tau_k)$ as the Laplace transform of $\rho(\omega)$, on $N$ points $\tau_k$ spaced by $\Delta \tau = 0.085 \text{ fm} = 0.43078 \text{ GeV}^{-1}$.
    The standard deviation of the noise is chosen as

    \[ \sigma(\tau_k) = d \, C_\text{orig}(\tau_k) \frac{\tau_k}{\Delta \tau}, \]
    where $d$ is a parameter which controls the noise level, identical to that of \cite{Asakawa:2000tr}. The mock data set is then generated as

    \[ C(\tau_k) = \mathcal{N}(\mu=C_\text{orig}(\tau_k), \sigma^2=\sigma\pqty{\tau_k}^2). \]
    These mock data sets were then inverted, ignoring $C(\tau_0)$, for various values of $N$ and $d$, using the method of \cref{sec:method}, to test the robustness of the method. No inversions without positivity constraints were performed, as significant positivity violations were observed in initial trials.  Here we used $N_\omega = 1000$ in the construction of $\boldsymbol{K}$. The results are shown in \cref{fig:toymodel}.
    Both more data-points, or less noise, are found to improve the quality of the reconstruction.
    Interestingly, more data-points do lead to a greater variance in the reconstructions, though their quality is improved with the number of data-points.
    A direct comparison with \cite[Figure 4]{Asakawa:2000tr} is complicated by the absence of uncertainties on that Figure, but the performance of the methods seems comparable using the naked eye.
	
	\section{Test: SU(2) $0^{++}$ Glueball}\label{sec:results_glueball}
	
	We depart from the Schwinger functions for the scalar SU(2) pure Yang-Mills glueballs as extracted from the data of \cite{Yamanaka:2019yek}, using lattice simulations for $\beta =$ 2.1, 2.2, 2.3, 2.4, 2.5. The Schwinger functions were computed using the raw data, that is, without any smearing applied. The lattice volumes, the number of configurations, the number of Schwinger function time slices $N$, and the uncertainties in each data set, are shown in \cref{tab:configs}.
	
	These simulations of \cite{Yamanaka:2019yek} relied on large ensembles of configurations, resulting in data sets with very low statistical uncertainties. This makes the results of the Tikhonov regularized inversion reliable, as seen in \cref{sec:results_meson}. Additionally, the number of Schwinger function time slices $N$ for these data sets is $7$, $9$, or $13$, meaning the variance in the inversion will be small, as was observed in \cref{sec:results_meson}.
	\begin{table}[tb]
		\centering
		\resizebox{\columnwidth}{!}{
			\begin{tabular}{l|l c r c c c} \hline
				$\beta$ & Volume & Configurations & $N$ & $\max(C(\tau)) / \expval{\sigma_C(\tau)}$ & $a \sqrt{\sigma}$ & $a[\Lambda^{-1}]$ \\
				\hline
				$2.1$ & $10^3 \times 12$ & 1,000,000 & 7 & $3.28 \times 10^{-4}$ & $0.608(16)$ & $0.356(27)$ \\
				$2.2$ & $12^4$           & 9,999,990 & 7 & $1.04 \times 10^{-4}$ & $0.467(10)$ & $0.273(20)$ \\
				$2.3$ & $14^3 \times 16$ & 4.100,000 & 9 & $1.38 \times 10^{-4}$ & $0.3687(22)$ & $0.216(15)$ \\
				$2.4$ & $16^3 \times 24$ & 2,030,000 & 13 & $1.55 \times 10^{-4}$ & $0.2660(21)$ & $0.156(11)$ \\
				$2.5$ & $20^3 \times 24$ & 520,000 & 13 & $3.04 \times 10^{-4}$ & $0.1881(28)$ & $0.110(8)$ \\
				\hline
			\end{tabular}
		}
		\caption{The glueball data sets of \cite{Yamanaka:2019yek}. Selection of information taken from \cite[Table~I, Table~II]{Yamanaka:2019yek}.}
		\label{tab:configs}
	\end{table}
	\begin{figure}[tb]
		\centering
		\begin{floatrow}
			\floatbox{figure}[.45\columnwidth][\FBheight][b]
			{\caption{$C(\tau)/C(\tau_1)$ for all data sets.
				}\label{fig:C_all}}
			{\includegraphics[width=0.45\textwidth,trim=0.1in 0.1in 0.1in 0.1in ,clip]{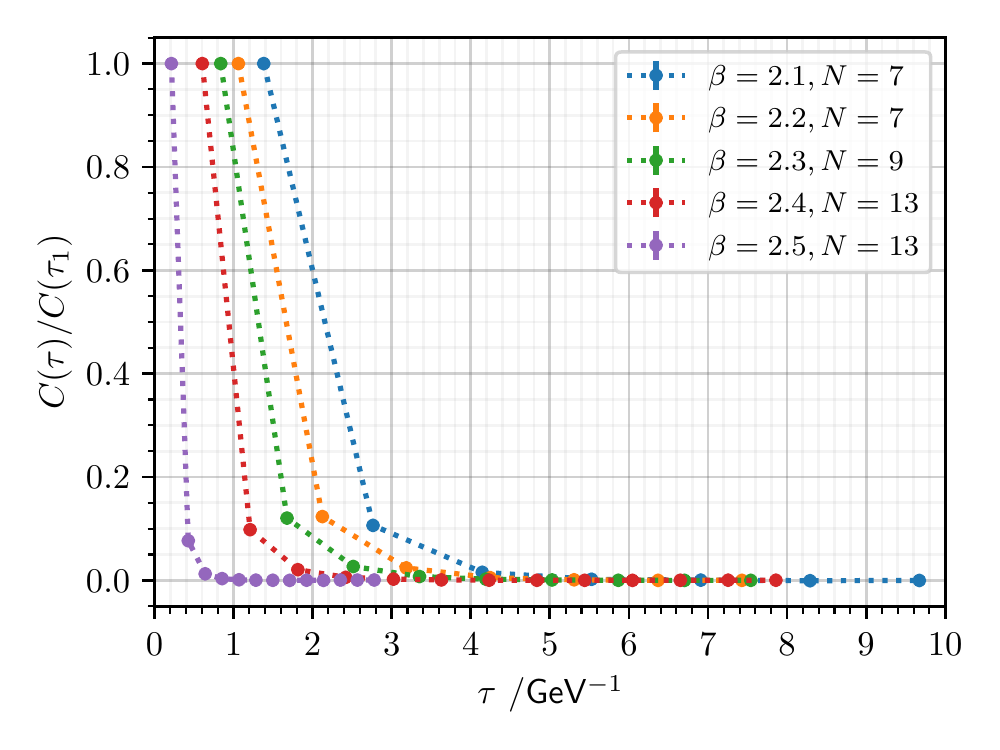}}
			\floatbox{figure}[.45\columnwidth][\FBheight][b]
			{\caption{$\rho(\omega)$ for all data sets, using the $ip$-method. The $y$-axis has been normalized to give the last peak an intensity of 1. Clearly positivity constraints should be imposed.
				}
				\label{fig:rho_all_ip}}
			{\includegraphics[width=0.45\textwidth,trim=0.1in 0.1in 0.1in 0.1in ,clip]{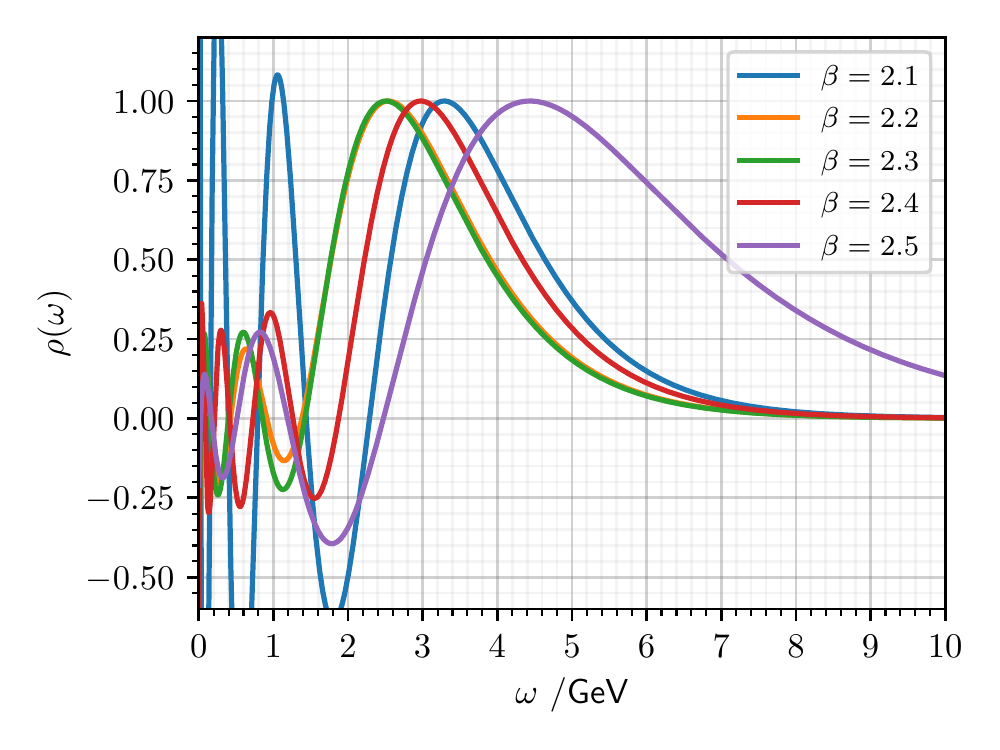}}
		\end{floatrow}
	\end{figure}
	
	The Schwinger functions for the various simulations can be seen in \cref{fig:C_all}, whereas \cref{fig:rho_all_ip} shows the spectral density functions as obtained without any constraints on $\rho(\omega)$, using the $ip$-method published in \cite{Dudal:2019gvn}.

\begin{figure}[htb]
		\centering
		\begin{floatrow}
			\floatbox{figure}[\FBwidth][\FBheight][t]
			{\caption{$\rho(\omega)$ for all datasets, subject to $\rho(\omega)~\geq~0$. The $y$-axis has been normalized to give the maximum an intensity of 1.}
				\label{fig:rho_all}}
			{\includegraphics[width=0.5\textwidth,trim=0.1in 0.1in 0.1in 0.1in ,clip]{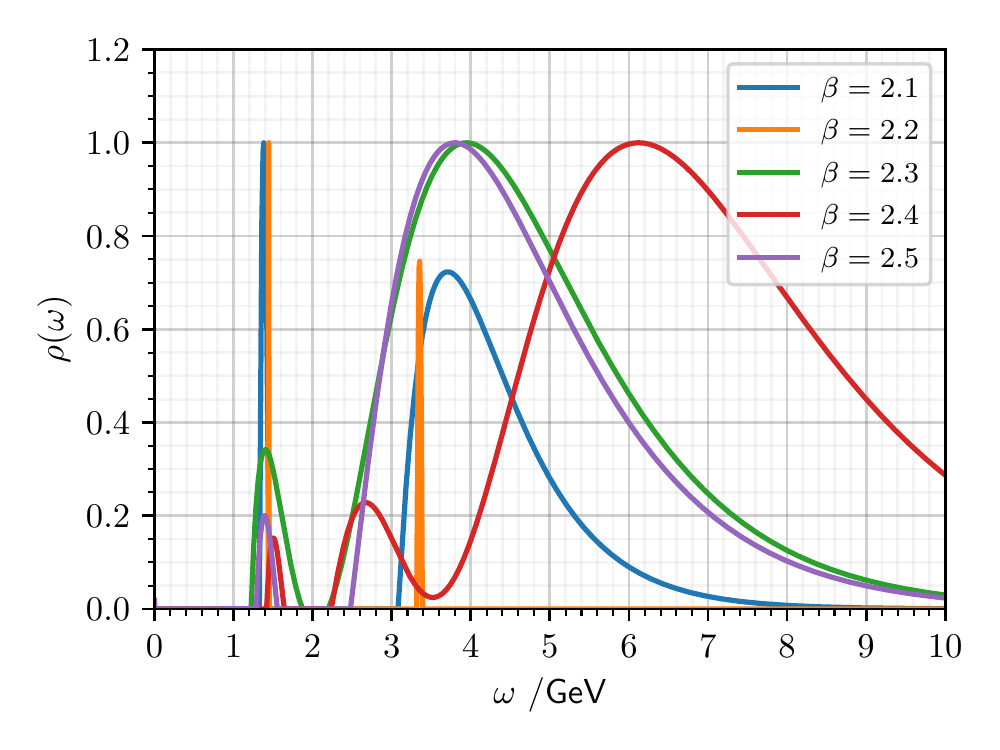}}
			\floatbox{table}[\FBwidth][\FBheight][t]
			{\caption{Maxima of \cref{fig:rho_all} in order of increasing $\omega$\SI{}{/\GeV}, with corresponding left and right Half Width at Half Maximum (HWHM).}\label{tab:states}}
			{
				\small
				\begin{tabular}{|l|l|}
					\hline
					$\beta$ & $\omega \pm \text{HWHM} \; / \text{GeV}$ \\
					\hline
					2.1 & $1.382 + 0.049 - 0.041$ \\
					& $3.712 + 1.048 - 0.486$ \\
					\hline
					2.2 & $1.444 \pm 0.003$ \\
					& $3.354 + 0.023 - 0.031$ \\
					\hline
					2.3 & $1.401 + 0.241 - 0.138$ \\
					& $3.941 + 1.929 - 1.106$ \\
					\hline
					2.4 & $1.505 + 0.085 - 0.064$ \\
					& $2.670 + 0.431 - 0.307$ \\
					& $6.118 + 2.644 - 1.520$ \\
					\hline
					2.5 & $1.392 + 0.103 - 0.079$ \\
					& $3.798 + 1.833 - 0.950$ \\
					\hline
				\end{tabular}
			}
		\end{floatrow}
	\end{figure}

	As \cref{fig:rho_all_ip} illustrates, there are significant positivity violations and rapid oscillations when positivity is not enforced,
	especially for small $\omega$, a tell-tale sign of overfitting. On the other hand, when positivity is imposed, the resulting spectral
	functions in \cref{fig:rho_all} all display a clear mass gap that correspond to a ground state mass of \SI{\sim 1.4}{ \GeV}. Moreover, the infrared oscillations are gone.
	
	The estimated (central value) mass values are given in \cref{tab:glueballmass}, which lists the $0^{++}$ ground state mass as obtained from the spectral density function, as well as those
	calculated by Yamanaka et al.~\cite{Yamanaka:2019yek}. For an error analysis, we refer to \cite{Dudal:2021gif}.
	
	\begin{table}[ht]
		\centering
		\begin{tabular}{l|lll|lll}
			& \multicolumn{3}{c}{Traditional \cite{Yamanaka:2019yek}} & \multicolumn{3}{c}{Spectral representation} \\
			\hline
			$\beta$ & $a m_\phi$ & $m_\phi[\Lambda]$  & $m_\phi$ /GeV & $a m_\phi$ & $m_\phi[\Lambda]$  & $m_\phi$ /GeV \\
			\hline
			2.1 & 1.853 & 5.21 & 1.34 & 1.910 & 5.364 & 1.382 \\
			2.2 & 1.517 & 5.55 & 1.43 & 1.532 & 5.613 & 1.444  \\
			2.3 & 1.241 & 5.75 & 1.48 & 1.174 & 5.436 & 1.401  \\
			2.4 & 0.924 & 5.93 & 1.53 & 0.910 & 5.832 & 1.505 \\
			2.5 & 0.696 & 6.32 & 1.63 & 0.595 & 5.408 & 1.392
		\end{tabular}
		\caption{The central values of the $0^{++}$ glueball masses as presented in Table 3 of \cite{Yamanaka:2019yek} compared with the spectral representation method. The physical units were calculated assuming $\sqrt{\sigma} = 0.44$ GeV.}
		\label{tab:glueballmass}
	\end{table}

	The values of $a \sqrt{\sigma}$ and $a[\Lambda^{-1}]$ are given in \cref{tab:configs}. A string tension $\sqrt{\sigma}$ of $\SI{0.44}{\GeV}$ was assumed to convert to physical units.
		
	The scalar glueball masses as obtained by \cite{Yamanaka:2019yek}, and those extracted using the spectral density method, agree quite well, despite the fact that we based ourselves on the unsmeared data, whilst the mass esimates from \cite{Yamanaka:2019yek} were obtained after smearing, which usually improves the ground state signal.
	
	In addition to the ground state, the spectra presented in \cref{fig:rho_all} also show hints of an excited state, though its position shows much greater variance with $\beta$ than that of the ground state. \Cref{tab:states} lists the $\omega / \text{GeV}$ values of all the local maxima in \cref{fig:rho_all}, and the left and right Half Width at Half Maximum (HWHM) values of each local maximum.

Except for $\beta = 2.4$, all conditions indicate a single excited state at $3.3-3.9$ \SI{}{\GeV}. The only exception is $\beta=2.4$, which has a second maximum at \SI{2.670(25)}{\GeV} and a third one at \SI{6.118(57)}{\GeV}. The relatively wide peaks for the excited states indicate that we might need a better signal, as there are in principle no decay channels open for these 1st excited states in pure gluodynamics, not even to the $0^{++}$ ground state.  Needless to say, one must be careful to interpret these values, as due to the finite lattice spacing, there is natural UV cut-off in the system. This will be discussed in more detail elsewhere.

	\section*{Acknowledgements}
	The authors are indebted to N.~Yamanaka, H.~Iida, A.~Nakamura and M.~Wakayama for sharing the Schwinger function estimates based on their data of \cite{Yamanaka:2019yek}.
	O.O.~was partly supported by the FCT – Funda\c{c}\~ao para a Ci\^encia e a Tecnologia, I.P., under project numbers UIDB/04564/2020 and UIDP/04564/2020, while the work of D.D.~and M.R.~was supported by KU Leuven IF project C14/16/067.

\bibliographystyle{unsrt}
\bibliography{biblio.bib}

\end{document}